\documentclass[openacc]{rstransa}

\usepackage{epsfig}
\usepackage{amsmath}
\usepackage{graphicx}
\usepackage[normalem]{ulem}
\usepackage[dvipsnames]{xcolor}

\newcommand{\diff}[1]{\mathrm{d}#1}

\begin{document}

\title{QCD Factorization and Quantum Mechanics}

\author{
C.~A.~Aidala$^{1}$ and T.~C.~Rogers$^{2}$}

\address{
$^1$Department of Physics, University of Michigan, Ann Arbor, MI 48109, USA \\
$^2$Department of Physics, Old Dominion University, Norfolk, Virginia 23529, USA \\
Jefferson Lab, Newport News, Virginia 23606, USA

}

\subject{QCD}


\corres{C.~A.~Aidala\\
\email{caidala@umich.edu}}

\begin{abstract}
It is unusual to find QCD factorization  explained in  the language  of  quantum  information  science.  However, we  will  discuss  how  the  issue  of factorization  and  its breaking in high-energy QCD processes relates to phenomena like decoherence and entanglement.  We will elaborate with several examples and explain them in terms familiar from basic quantum mechanics and quantum information science.
\end{abstract}


\begin{fmtext}
\section{Introduction}
\label{e.intro}

The quark-gluon parton model is the conceptual basis upon which most scattering experiments seek to study the constituent quark and gluon structure of hadrons.
The parton model gets its justification in perturbative QCD (pQCD) and in particular from the QCD factorization theorems. The basic partonic picture -- see, for example, Feynman's original formulation of it in Ref.~\cite{Feynman:1972} -- is an essentially semi-classical picture of scattering between hadron constituents, wherein specific well-defined events occur over specific ranges of spacetime and in a specific sequence. Indeed, decoherence is one 
of the main ingredients of the parton model as it is normally taught~\cite{EllisStirlingWebber}. The purpose of this article is 
to highlight the ways that the aims of QCD factorization derivation often overlap with topics normally considered the domain of quantum information theory and interpretations of quantum mechanics~\cite{Laloe:2002cay}. 

It is useful to start by recalling, in picture form, the basic description of the parton model for deep inelastic scattering (DIS). It proceeds through the stages shown in Fig.~\ref{DIS_diagram}. First, an electron and a proton approach one another in the center of mass frame at high velocity (Fig.~\ref{DIS_diagram}-A). The proton is assumed to be a cluster of small constituents. 

\end{fmtext}


\maketitle

\begin{figure}[!h]
\centering\includegraphics[width=4.75in]{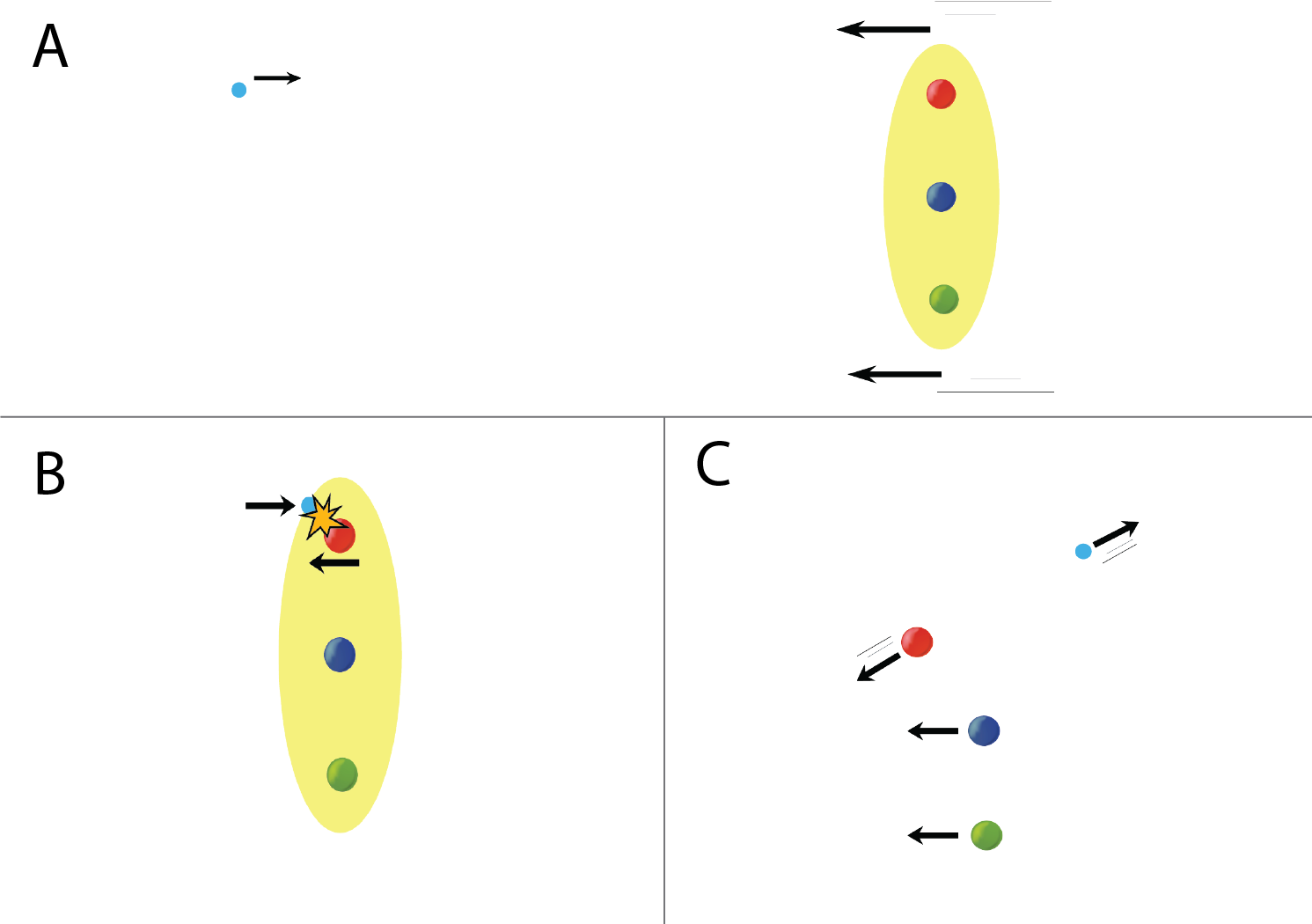}
\caption{The sequence of distinct steps in a typical description of DIS. A) An electron (left) and hadron (right) approach. B) The hard collision between an electron and a (red) parton. C) A recoiling electron-parton pair and a final state of free partons.}
\label{DIS_diagram}
\end{figure}

Next, the electron collides with one and only one of the parton constituent and scatters at wide angle (Fig.~\ref{DIS_diagram}-B). The struck quark itself recoils at wide angle. All the remnant partons continue to propagate nearly freely and at high speed until, long after the collision, they are separated by large distances (Fig.~\ref{DIS_diagram}-C). Pictures like these suggest essentially classical probabilities. For the cases where only the scattered lepton is observed, there is a sum over all ways that the final state 
partons can interact.  That formula is simply 
\begin{equation}
\label{DIS}
\diff{\sigma} = \mathcal{H} \otimes f(x) \, ,
\end{equation}
where $\sigma$ represents a cross section, $\mathcal{H}$ represents the likelihood for the electron-quark collision, and  $f(x)$ is a probability density (or number density) for the quark with momentum fraction $x$ to be found in the incoming proton. The incoming and outgoing partons are 
treated as asymptotic states, even though it is known that 
the true asymptotic states in QCD are hadrons rather than quarks or gluons.
It is a very classical description that is familiar from textbooks and lectures. 

\begin{figure}[!h]
\centering\includegraphics[width=4.0in]{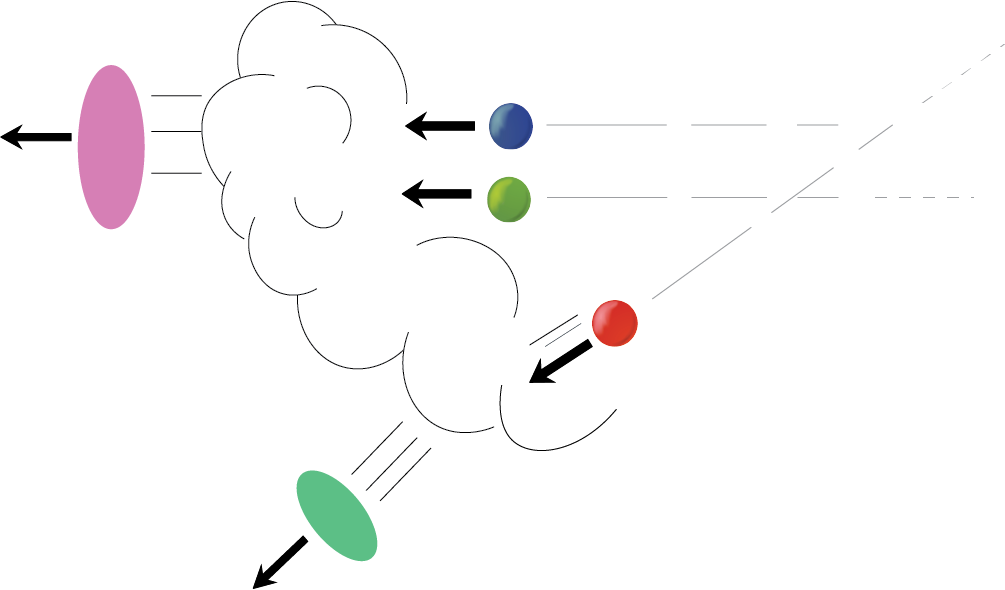}
\caption{A hadronization process in semi-inclusive DIS, adding a final step after panel C in Fig.~\ref{DIS_diagram}.}
\label{hadronize_diagram}
\end{figure}

The situation is more interesting if one observes the actual final state particles (which in QCD are always hadrons) and places restrictions on which ones are to be counted. For example, one might count the number of high momentum pions (still summing over everything else). Events without pions in a designated range of momentum are excluded from this cross section. Then, a typical picture of the end stage of the process is shown in Fig.~\ref{hadronize_diagram}. The electron and hadron still approach one another as in~\ref{DIS_diagram}-A, and a parton is again struck hard by the electron, as was already illustrated in \ref{DIS_diagram}-B, but now one recognizes that, a long time and distance after the collision, the struck quark undergoes a transformation (hadronization) into a constituent of one of the counted pions. (Of course, all of the other partons also hadronize as well.) The details of the hadronization mechanism are still only vaguely understood, and in Fig.~\ref{hadronize_diagram} it is represented by the puffy clouds. The turquoise oval is the observed final state hadron while the lavender oval represents whatever other unobserved final state particles might be present. 

In this picture, the probability that the initial red parton arrives with fractional momentum $x$ is determined by processes associated with proton structure that occur a very long time and distance before the collision (the far right of \ref{DIS_diagram}-A).  By contrast, the probability for the final red parton hadronizing with momentum fraction $z$ into a final pion is dictated by processes a long time and distance after the collision, and a large distance from the other hadronization processes represented by the lavender oval. The two mechanisms just described are associated with the parton densities $f(x)$ and the  fragmentation functions $D(z)$, respectively. The probability $\mathcal{H}$ for the electron-parton collision involves only very short time and distance scales.

The large time and distance scales between the initial hadron structure mechanisms embodied by $f(x)$, the partonic collision itself, and the final state parton hadronization embodied by $D(z)$ suggests that it is reasonable to view each as independent and uncorrelated. In particular, $D(z)$ is an intrinsic and universal property of a particular type of high-momentum hadronizing parton, independent of the details of the larger process that it is a part of. In Fig.~\ref{hadronize_diagram}, this means the lower part of the picture that involves the red parton and the turquoise oval are independent of the upper part that involves the remaining partons and the lavender oval.

If this classically probabilistic picture holds, then the likelihood for the process (usually presented in the form of a cross section $\diff{\sigma}$) is just the product of each of the uncorrelated component probabilities: 
\begin{equation}
\label{SIDIS}
\diff{\sigma} = \mathcal{H} \otimes f(x) \otimes D(z) \, .
\end{equation}
In practice, QCD factorization formulas like 
Eq.~\eqref{DIS} and Eq.~\eqref{SIDIS} do not involve simple products, but rather convolution integrals over shared momentum fractions, and this is represented by the ``$\otimes$'' symbol. 

As might be expected, such simplistic pictures as these come with many caveats, and they are only approximate when they hold at all. But they provide a useful starting point for more detailed studies, and are ubiquitous in investigations of partonic structure by way of scattering experiments. 

The practical usefulness of factorization formulas like these lies in the fact that QCD is asymptotically free over short time and distance scales, so that it is possible to calculate $\mathcal{H}$ (which describes the actual collision with the quark) using ordinary small coupling perturbation theory techniques. The other pieces of the formula, $f(x)$ and $D(z)$, are to be viewed as intrinsic to the incoming proton and hadronizing parton, respectively. These nonperturbative functions are independent of the particular physical process and can be constrained by experimental data. 

\begin{figure}[!h]
\centering\includegraphics[width=3.9in]{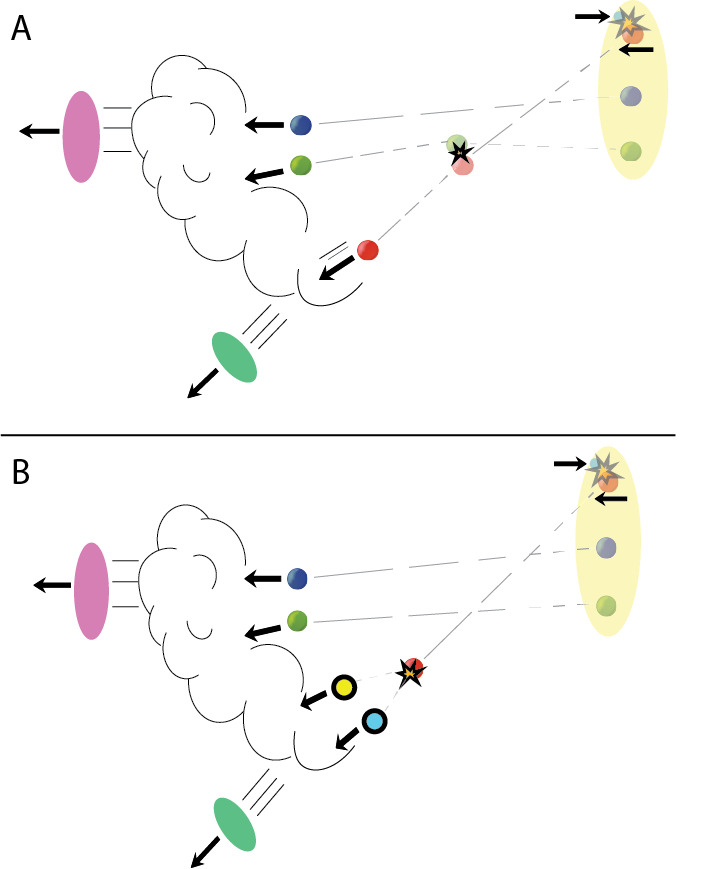}
\caption{Alternatives to Fig.~\ref{hadronize_diagram} for the final state description of semi-inclusive deep inelastic scattering.}
\label{hadronize_complex}
\end{figure}

The semi-classical nature of the factorization formulas above suggests that their derivations in real QCD are ultimately arguments about decoherence between processes happening over widely separate time and distance scales. And departures from the simplest factorized pictures typically involve some form of failure of decoherence. Since the original fully inclusive DIS process in Eq.~\eqref{DIS} involves a sum over a large number of final states, decoherence washes out the large-distance interference effects that might otherwise spoil the simple treatment of the final configuration as a collection of free quarks and gluons. In the semi-inclusive case in Eq.~\eqref{SIDIS}, the number of final state interactions has been reduced relative to the fully inclusive case in Fig.~\ref{DIS_diagram}, along with the opportunities for decoherence to occur, so more work is needed there to demonstrate factorization. This is clear from the number of possible alternatives to Fig.~\ref{hadronize_diagram} as descriptions of the final state. Some examples are shown in Fig.~\ref{hadronize_complex}. In panel A, the struck parton interacts with one of the other partons in the final state before hadronizing. In panel B the struck parton splits into two different partons, implying that the outgoing red parton is not actually an asymptotic state. Of course, one must consider that the physical pictures in Fig.~\ref{hadronize_diagram} and Fig.~\ref{hadronize_complex}A-B can all interfere with one another quantum mechanically, so in principle none can be said to be the ``true'' picture. Other contributions than these can entangle initial and final states, complicating the situation even further. If they are localized to short spacetime regions, these types of contributions might be accommodated through perturbative corrections within $\mathcal{H}$, but the possibility that they might involve large spacetime regions threatens to spoil the general partonic picture itself.  
It is a non-trivial result, therefore, that a factorization in the form of Eq.~\eqref{SIDIS} nevertheless holds true, though there are some cases (e.g., when the observed final state depends on intrinsic transverse momentum) that interesting departures from naive expectations do arise. 

We began this discussion with the very course-grained inclusive process in Eq.~\eqref{DIS} and noted the potential for a breakdown in decoherence effects if we moved to the more detailed semi-inclusive process in Eq.~\eqref{SIDIS}. Of course, one may continue this trend by considering even more details and counting, for example, the numbers of more complex events multihadron final states. If this gets continued indefinitely, one eventually expects the decoherence that allows for a factorized picture to break down. Where and exactly how that happens is an interesting open question.

\section{Spin and entanglement}
\label{e.spin_entanglement}

Entanglement phenomena play an especially central role in QCD factorization theorems when they are generalized to include processes involving spin and/or intrinsic transverse momenta. A prime example is
the Collins effect, which we will describe now. 

A physical process that is very useful for describing the effect is semi-inclusive  electron-positron annihilation (SIA) with a pair of 
nearly back-to-back hadrons $h_A$ and $h_B$ produced inclusively in the final state:
\begin{equation}
e^+ + e^- \to \gamma^\ast \to h_A + h_B + X \, .
\end{equation}
The electron-positron pair creates a highly virtual photon $\gamma^\ast$ that splits into a quark-antiquark pair. The quark and antiquark each hadronize and become constituents of the observed hadrons $h_A$ or $h_B$ in a manner similar to what was discussed above for semi-inclusive deep inelastic scattering. 
 The $X$ indicates that all other final states are unobserved.  
 The inclusive cross section is a measure of the total number of occurrences of hadron pairs of types $h_A$ and $h_B$ in nearly back-to-back configurations.

\begin{figure}[!h]
\centering\includegraphics[width=4.5in]{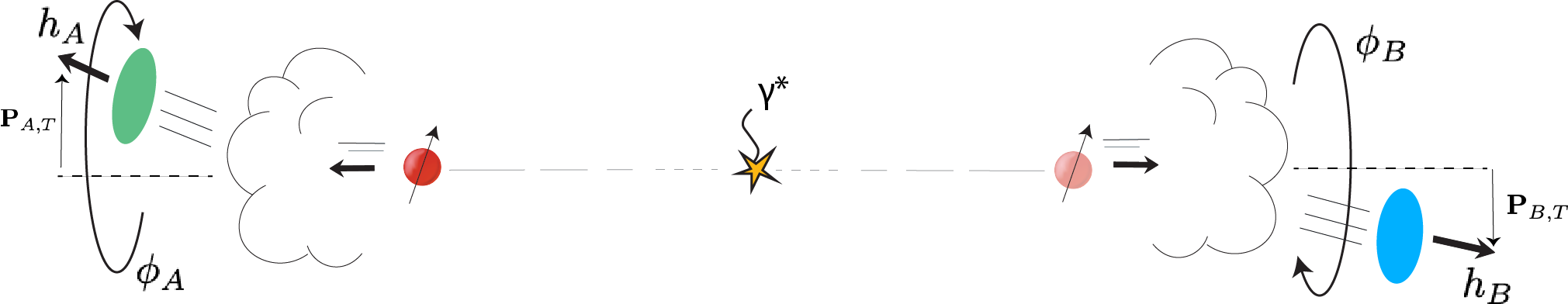}
\caption{A schematic of the products of an $e^+ e^-$ collision showing the virtual photon in the center. A quark (red) and anti-quark (faded red) move apart from each other in a back-to-back configuration. The cloud puffs represent the hadronization process. Hadrons other than $h_A$ and $h_B$ are also produced but are not shown in the figure. The fact that the pair's spins can remain entangled is represented by the vertical arrows.}
\label{SIApicture}
\end{figure}
Figure~\ref{SIApicture} is a visualization of the last stages of the process   
in a reference frame where the quark-antiquark pair has zero total momentum. It suggests a sequence of events very analogous to what we discussed above with SIDIS. First, there is a hard collision represented by 
the central star, and this can be handled with very high precision using small coupling techniques. Next, the quark and anti-quark move away from each other until they attain a wide spatial separation. At this point, they independently transform into constituents of the observed hadrons $h_A$ and $h_B$. Since it is an inclusive process, there are potentially many other final state particles that go uncounted. These uncounted particles are analogous to the lavender oval in Fig.~\ref{hadronize_diagram}, but we leave this implicit in Fig.~\ref{SIApicture} to keep the diagram simple. 
QCD factorization theorems confirm that an equation analogous to Eq.~\eqref{SIDIS} also holds for this process:
\begin{equation}
\label{SIA}
\diff{\sigma} = \mathcal{H} \otimes D_A(z_A,{\bf P}_{A,T}) \otimes D_B(z_B,{\bf P}_{B,T}) \, .
\end{equation}
Now there are two fragmentation functions $D_A$ and $D_B$ representing the clouds on the far right and left sides of Fig.~\ref{SIApicture} where the transformation of the (anti)quark/parton into a hadron constituent occurs. Note, however, that this time we have explicitly included dependence on a small transverse component of momentum for each of the final hadrons. These two-component vectors describe the hadrons' deviation from a back-to-back configuration. They are labeled  ${\bf P}_{A,T}$ and ${\bf P}_{B,T}$ in Eq.~\eqref{SIA}, and in this form the equation is the transverse momentum dependent (TMD) version of the factorization theorem. 

Now each of the ${\bf P}_{A/B,T}$ vectors has an azimuthal angle $\phi_{A/B}$ that determines its orientation around the central axis. In each of the hadronization clouds, there is nothing to pick out a particular azimuthal angle, the partons and hadrons are all unpolarized, and the clouds in 
Fig.~\ref{SIApicture} have a large spacelike separation. So, in a purely classical picture, we would not expect any azimuthal dependence in the cross section. 

However, the quark and antiquark, despite being unpolarized, nevertheless have entangled spins, and this results in a measurable azimuthal dependence. The situation is very reminiscent of the Einstein-Podolsky-Rosen scenario~\cite{Einstein:1935rr}, but with the process of hadronization taking the place of separate, distant measurements.  The entangled spins are allowed in the TMD factorization theorem for this process, with fragmentation functions each carrying transverse spin indices. The resulting azimuthal dependence  is called the Collins effect~\cite{Collins:1992kk}. 

This example illustrates how factorization theorems lose some of their naively classical features as one examines the finer details involved in a process, such as small kicks of transverse momentum. The inverse is true if there is an average over the final state details. In the single inclusive annihilation (SIA) example above, integrating over all the ${\bf P}_{A/B,T}$ eliminates sensitivity to the type of entanglement effects just described. 

\section{Factorization breaking}

\begin{figure}[!h]
\centering\includegraphics[width=4.75in]{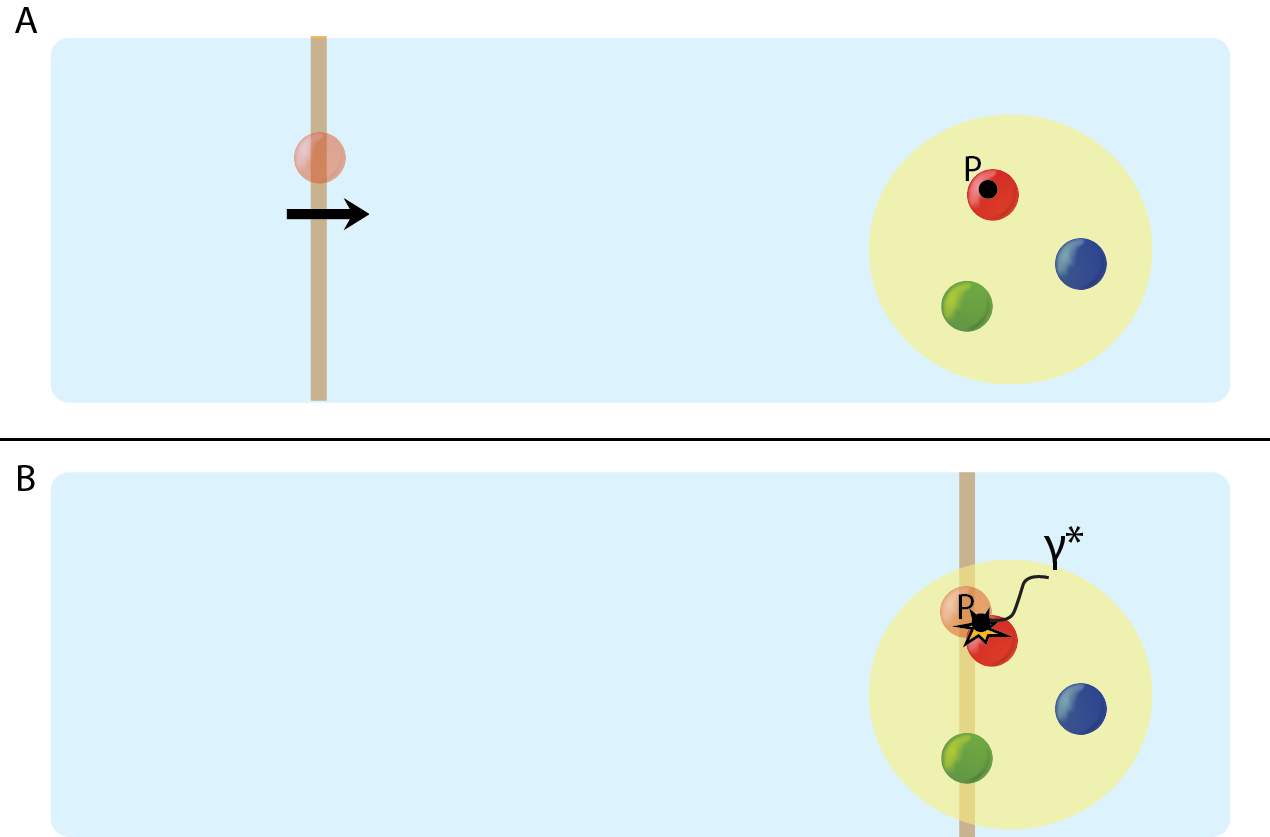}
\caption{A schematic illustration of the Basu-Ramalho-Sterman picture in the case of Drell-Yan scattering. In A, a red anti-quark approaches a stationary proton on the right at high speed. (It is presumably associated with its own hadron, not shown.) $P$ is the point of observation in the proton. The pale blue field represents the vector potential spreading out through space, while the vertical orange band is the ``pulse'' force associated with the incoming antiquark. In B, the antiquark annihilates with the quark. It is only when the orange overlaps with both the red and faded-red partons that the two may exert forces on one another.}
\label{DY_diagram}
\end{figure}

A number of phenomena can break classically inspired pictures of factorization
in more significant ways than have been described so far. 

In the descriptions of it above, QCD factorization describes a type of decoherence between processes that occur at widely separated locations in space and time. For it to be possible, of course, there must be a hierarchy of scales in the process. In the SIDIS and SIA examples above, that was provided by the virtuality $Q^2$ of the exchanged/annihilation photon, and each process involves a range of scales from small distance scales of the order of the resolution scale $\lesssim 1/Q$ to the large distances involved in hadronization. In QCD, this hierarchy is especially important because asymptotic freedom over short distance allows the basic partonic process to be calculated directly in terms of propagating quark and gluon degrees of freedom. As such, the most basic way that factorization can break down is when the relevant hard scales become too low. Then there is no distinction between small and large distance scales, and the QCD coupling $\alpha_s(Q)$ is no longer small enough for one to speak of 
``perturbative'' or ``nonperturbative'' parts. 

But there are other more interesting ways that factorization theorems (or specific aspects of factorization theorems) can potentially be violated, even when hard scales are large. They are frequently associated with gauge invariance; the 
necessity to include non-trivial Wilson line 
operators in the definitions of parton 
distribution and fragmentation functions tends to introduce effects that violate the most naive number density interpretations. 

 Basu, Ramalho, and Sterman~\cite{Basu:1984ba} composed a nice classical analogy to help with interpreting what happens with gauge fields (like QCD) in those hadron-hadron collisions where factorization theorems hold. It will be helpful to reproduce a description of it here.\footnote{Reference~\cite{Basu:1984ba} discusses hadron-hadron collisions more generally, but for definiteness we will specialize to the Drell-Yan process, wherein two hadrons collide and an antiquark from one annihilates with a quark from the other to produce a high-mass lepton-antilepton pair.} 
 They note that, compared with simpler hard processes like DIS,  there is a greater possibility for factorization breaking effects in hadron-hadron collisions because the color field of one hadron might alter the structure of the other hadron before their respective constituent partons have an opportunity to collide. If such interactions are large, then they would break factorization very much. 
 
 To build an intuition, Ref.~\cite{Basu:1984ba} considers the gauge field of an incoming point-like quark from one of the hadrons as viewed from the rest frame of the other. (See Fig.~\ref{DY_diagram}). They note that if the gauge field were a \emph{classical}, abelian field like electromagetism, then its vector potential at a coordinate $P = ({\bf x}_T, x^z)$ would follow from standard electromagnetism textbooks:
 \begin{align}
 A^0(P) &{}= \frac{q \gamma}{\sqrt{{\bf x}_T^2 + \gamma^2 (\beta t - x^z)^2}} \, , \\
 A^z(P) &{}= \frac{-q \gamma \beta}{\sqrt{{\bf x}_T^2 + \gamma^2 (\beta t - x^z)^2}} \, , \\
 A^x(P) &{} = A^y(P) = 0 \, ,
 \end{align}
 with $q$ being the charge of the quark and $\gamma$ and $\beta$ being the usual special relativity symbols. In the limit that the center-of-mass $\sqrt{s}$ of the collision approaches infinity, the time and $z$ components of the vector potential do not vanish for any ${\bf x}$ or time $t$, so that one might worry that the quark in the incoming hadron influences the structure of the other hadron long before the collision of partons actually takes place. This would constitute a violation of factorization. However, the vector potential takes the form 
 \begin{equation}
 \lim_{s \to \infty} A^\mu(P) = q \partial^\mu \ln (\beta t - x^z) \, .
 \end{equation}
 That is, it is a pure gauge potential for all times and places except where $x^z = \beta t$. The quark coming in from the left exerts no significant forces at the observation point in the target hadron on the right except right at the instant when it is passing through it. Indeed, a direct computation of the $z$-component of the electric field gives
 \begin{equation}
 E^z(x^z \neq \beta t) \sim 1/s^2 \, .
 \end{equation}
Therefore, any influence the field from the incoming quark might have on the right-side hadron before the collision is strongly power suppressed in the high energy limit. There is no chance of the hadrons significantly influencing one another before the hard collision of the partons, and so a factorized picture appears to hold. 

In the standard factorization theorems for inclusive processes, one is forced to deal with an infinitely large number of 
longitudinally polarized gluon exchanges that appear to threaten factorization graph-by-graph. A significant amount of the effort in deriving a factorization theorem for a process like Drell-Yan scattering in real QCD amounts to using arguments based on gauge invariance to show how a cascade of cancellations removes any resulting factorization breaking effects from these ``extra'' gluons in specific physical observables. The picture from Ref.~\cite{Basu:1984ba} leads to an interpretation of this as confirmation that transitioning to a nonabelian quantum field theory preserves the normal expectation from a classical analogy. 

Formally, the longitudinally polarized gluons simply get absorbed into the Wilson line operators of separate parton densities, where they appear to play the purely formalistic role of ensuring gauge invariance for parton densities. The Wilson lines are rather abstract objects, but they essentially just account for the the fact that a colliding parton must pass through a background field on its way in or out of a hard collision, similar to the description in Fig.~\ref{DY_diagram}. The phases accumulated along those paths need to be accounted for to preserve formal gauge invariance in parton densities. But at least in the simplest and most highly inclusive cases, they do not impact the general overall partonic picture. Indeed, specializing to the light-cone gauge reduces the Wilson line in a parton density to a unity operator, and the parton density gets an interpretation very close to that of an actual number density. 

In quantum mechanics, a gauge field can have an influence on a particle even when it is confined to a region where the actual force vanishes, as famously illustrated by the Aharonov-Bohm effect~\cite{Aharonov:1959fk}. Therefore, it is not as trivial as it might initially seem to prove that classical expectations from the Basu-Ramalho-Sterman or similar pictures carry over to QCD for other observables. As in the SIA, SIDIS and Drell-Yan cases discussed above, factorization for many standard processes relies on the fact that the observables are very inclusive, with enough final state averaging to give the necessary cancellations. If one constructs observabes with greater sensitivity to details, interesting effects can modify or even break classical expectations. 

One place where that happens is when the effects of a small intrinsic transverse momentum are studied. (We already encountered one example of this in the discussion of Eq.~\eqref{SIA}.) For instance, we might ask how an intrinsic transverse momentum of a bound state quark in a proton gets influenced if the proton is transversely polarized. This is called a Sivers effect~\cite{Sivers:1989cc}. It turns out that if the proton is regarded entirely in isolation, without reference to any process, then an argument based on CP invariance shows that such a correlation must vanish~\cite{Collins:1992kk}. While factorization theorems nevertheless hold for processes like SIDIS and the Drell-Yan process where such an effect might be observed, it turns out that the structure of the Wilson line operator in the parton densities is crucial. A different structure arises 
if it is an outgoing struck quark that passes through the gauge field \emph{after} the collision, as in the 
SIDIS example of Fig.\ref{hadronize_diagram}, or if it is in incoming quark passing through on the way \emph{into} the collision as in Fig.~\ref{DY_diagram}. The result is a nonzero Sivers effect, but with a sign that flips between these two different scenarios. 
\begin{figure}[!h]
\centering\includegraphics[width=2.5in]{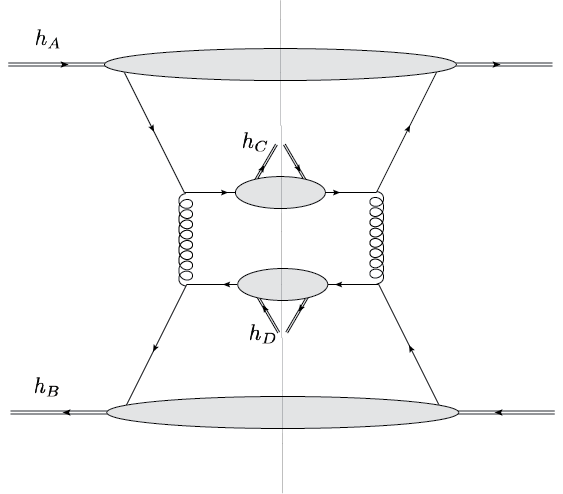}
\caption{A typical Feynman diagram for the squared amplitude of the process in Eq.~\eqref{hadron_proc}.}
\label{hadronsdiagram}
\end{figure}

One might infer from the Sivers example that factorization always holds when there is a hard scale, with the only caveat being that the Wilson line operator becomes process dependent. Indeed, it is easy to check that this appears to be the case for \emph{abelian} gauge fields. However, it turns out that separate Wilson lines are not possible in nonabelian cases like QCD. One of the simplest examples is a high energy collision of two hadrons with a hard production of a nearly back-to-back hadron pair produced inclusively in the final state:
\begin{equation}
\label{hadron_proc}
h_A + h_B \to h_C + h_D + X \, .
\end{equation}
As in the SIA example, we measure slight deviations from back-to-back, so some form of transverse momentum dependent factorization with transverse momentum dependent parton densities is needed.
The natural classical expectation, based on the rest of the discussion in this paper, is that a factorization formula emerges similar to Eq.~\eqref{SIA}, but with combinations of parton distribution and fragmentation functions for both the initial and final states~\cite{Bomhof:2006dp}:
\begin{equation}
\label{hadronfact}
\diff{\sigma} \stackrel{??}{=} \mathcal{H} \otimes f_A(x_A,{\bf P}_{A,T}) \otimes f_B(x_B,{\bf P}_{B,T}) \otimes D_C(z_C,{\bf P}_{C,T}) \otimes D_D(z_D,{\bf P}_{D,T}) \, ,
\end{equation}
but possibly with nontrivial and non-universal Wilson line operators for each of the four correlation functions. Each Wilson line represents an accumulation of phases as a parton passes in to or out of the hard collision. 

\begin{figure}[!h]
\centering\includegraphics[width=3.0in]{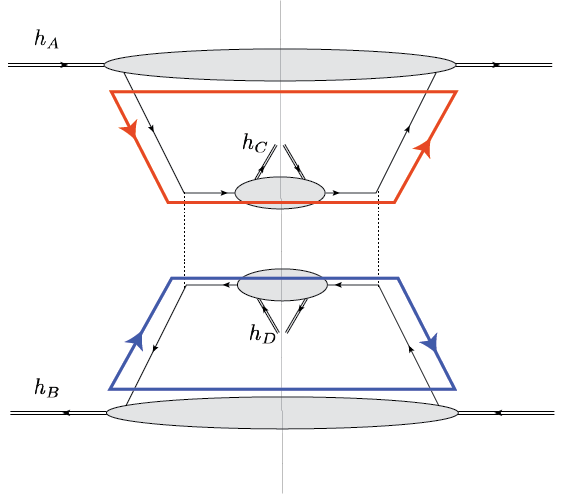}
\caption{A Feynman graph showing a tree-level contribution to the process in Eq.~\eqref{hadron_proc}. This graph is like Fig.~\ref{hadronsdiagram}, but the hard collision now involves the exchange of a hard colorless scalar (dotted) particle. Thus, at tree-level no color flows between the top and bottom of the graph. The red and blue colored loops show the flow of color charge in each part of the diagram.}
\label{fact_break1}
\end{figure}

\begin{figure}[!h]
\centering\includegraphics[width=3.5in]{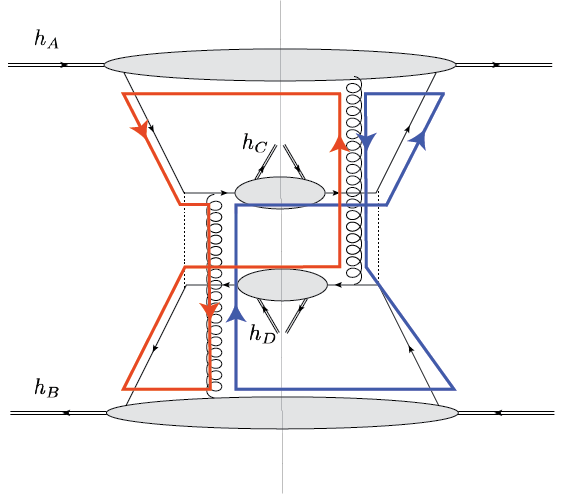}
\caption{Two longitudinally polarized infrared gluons, one collinear to hadron A and one collinear to hadron B. Contributions to scattering cross sections from graphs like these are generally non-zero. The red and blue lines show one possible path of color flow through the graph. Note that non-singlet color is carried by the gluons.}
\label{fact_break2}
\end{figure}

\begin{figure}[!h]
\centering\includegraphics[width=4.25in]{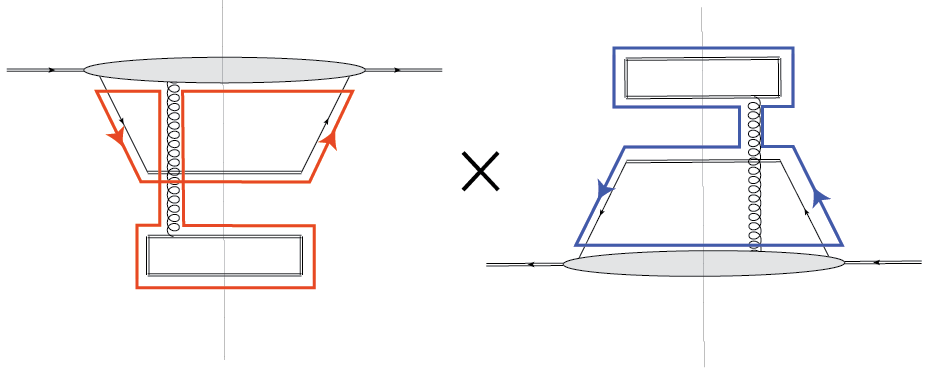}
\caption{For Fig.~\ref{fact_break2} to factorize as in Eq.~\eqref{hadronfact}, it must be possible to re-express the color flow as a product of two color loops associated with Wilson loops, as shown here. But each factor would involve an unphysical color singlet gluon exchange, so both factors must vanish. Meanwhile, the original contribution in Fig.~\ref{fact_break2} is non-vanishing. Therefore, the product above is not equal to Fig.~\ref{fact_break2} and the factorization fails.}
\label{wilson_lines}
\end{figure}

The exact reason that a formula like Eq.~\eqref{hadronfact} does not follow as in a normal factorization derivation is somewhat technical, and beyond the scope of what can be included in a short summary. However, the visualization tools in Figs.~\ref{fact_break1},~\ref{fact_break2} and \ref{wilson_lines} will help to convey the basic obstacle to a proof. It is a summary of a more detailed argument provided in~\cite{Rogers:2010dm}.

What Fig.~\ref{fact_break1} shows is a simpler version of a hard collision between quarks from separate hadrons than Fig.~\ref{hadronsdiagram}. We consider the case where a colorless particle is exchanged in the $t$-channel, so that at tree-level no color flows between the two hadrons. 
Next, we consider two of the type of longitudinally polarized infrared gluons that ordinarily would generate the Wilson lines that go into definitions for parton densities. A nontrivial flow of color charge through the two hadrons is possible, given the simultaneous presence of multiple gluons, as shown in Fig.~\ref{fact_break2}. The result is a nonvanishing contribution to physical observables. 

If a factorized expression were possible, then by definition it would be possible for each of those longitudinally polarized gluon attachments to be re-expressed as contributions to a separate Wilson line in a separate TMD parton distribution function. But then the only possible Wilson line structures that can arise from configurations like Fig.~\ref{fact_break2} are Wilson loops. However, these have to give a vanishing contribution since they involve exchange of a color singlet gluon, which is forbidden in QCD (see Fig.~\ref{wilson_lines}). Thus, there is an inconsistency between the conjectured factorization in Eq.~\eqref{hadronfact} and the flow of color charge that it would have to contain in order to be valid. One must conclude that, despite expectations, in general such an equation does not hold. The Wilson line structures cannot be associated with separate hadrons, but only with simultaneous combinations of several hadrons. Said differently, the factors accumulated as the partons propagate into and out of the interaction point are not simple products of the phase factors from parton A in hadron B and parton B in hadron A.  

\section{Conclusions}

In this article, we have remarked upon several places where the problem of 
QCD factorization overlaps with topics that are more commonly found in 
discussions about the fundamentals of quantum mechanics. The original factorization pictures formulated by Feynman and others are heavily motivated by classical intuition. Many of their features carry over to the quantum field theory case, provided that the physical processes under consideration average over the many interactions possible across large distance scales. We have emphasized the analogy between this and standard descriptions of quantum mechanical decoherence, and we have seen that the decoherence in QCD is necessary to give meaning to questions about the partonic structure of hadrons. 

Some of the examples we gave above are suggestive of ways to test or probe quantum mechanical features in QCD. We noted, for example, the entanglement exhibited by the Collins effect in Sec.~\ref{e.spin_entanglement}, and the similarity to the scenario of the EPR theorem. It is plausible that, with care, measurements could be formulated to test Bell inequalities. We note some recent proposals to test Bell inequalities in higher energy collisions~\cite{Barr:2021zcp,Takubo:2021sdk}.

\enlargethispage{20pt}

\funding{C.A.A. was supported by the U.S. Department of Energy, Office of Science, Office of Nuclear Physics, under Award Number DE-SC0013393, and by the National Science Foundation, under Award Number 2012926.  T.R. was supported by the U.S. Department of Energy, Office of Science, Office of Nuclear Physics, under Award Number DE-SC0018106.  This work was also supported by the DOE Contract No. DE-AC05-06OR23177, under which Jefferson Science Associates, LLC operates Jefferson Lab.}


\bibliographystyle{unsrt}
\bibliography{QCD_quantum}

\begin{thebibliography}{10}

\bibitem{Feynman:1972}
R.~P. Feynman.
\newblock {\em {P}hoton-{H}adron {I}nteractions}.
\newblock Benjamin, Reading, MA, 1972.

\bibitem{EllisStirlingWebber}
R.~Keith Ellis, W.~James Stirling, and B.~R. Webber.
\newblock {\em {QCD and Collider Physics}}, volume~8.
\newblock Cambridge University Press, 1996.

\bibitem{Laloe:2002cay}
Franck Laloe.
\newblock {Do we really understand quantum mechanics?}
\newblock {\em Am. J. Phys.}, 69:655--701, 2001.

\bibitem{Einstein:1935rr}
Albert Einstein, Boris Podolsky, and Nathan Rosen.
\newblock {Can quantum mechanical description of physical reality be considered
  complete?}
\newblock {\em Phys. Rev.}, 47:777--780, 1935.

\bibitem{Collins:1992kk}
John~C. Collins.
\newblock {Fragmentation of transversely polarized quarks probed in transverse
  momentum distributions}.
\newblock {\em Nucl.Phys.}, B396:161--182, 1993.

\bibitem{Basu:1984ba}
Rahul Basu, Anibal~J. Ramalho, and George~F. Sterman.
\newblock {Factorization at Higher Twist in Hadron - Hadron Scattering}.
\newblock {\em Nucl. Phys. B}, 244:221--246, 1984.

\bibitem{Aharonov:1959fk}
Y.~Aharonov and D.~Bohm.
\newblock {Significance of electromagnetic potentials in the quantum theory}.
\newblock {\em Phys.Rev.}, 115:485--491, 1959.

\bibitem{Sivers:1989cc}
Dennis~W. Sivers.
\newblock {Single Spin Production Asymmetries from the Hard Scattering of
  Point-Like Constituents}.
\newblock {\em Phys.Rev.}, D41:83, 1990.

\bibitem{Bomhof:2006dp}
C.~J. Bomhof, P.~J. Mulders, and F.~Pijlman.
\newblock {The Construction of gauge-links in arbitrary hard processes}.
\newblock {\em Eur. Phys. J. C}, 47:147--162, 2006.

\bibitem{Rogers:2010dm}
Ted~C. Rogers and Piet~J. Mulders.
\newblock {No Generalized TMD-Factorization in Hadro-Production of High
  Transverse Momentum Hadrons}.
\newblock {\em Phys.Rev.}, D81:094006, 2010.

\bibitem{Barr:2021zcp}
Alan Barr.
\newblock {Testing Bell inequalities in Higgs boson decays}.
\newblock 6 2021.
\newblock arXiv:2106.01377.

\bibitem{Takubo:2021sdk}
Yosuke Takubo, Tsubasa Ichikawa, Satoshi Higashino, Yuichiro Mori, Kunihiro
  Nagano, and Izumi Tsutsui.
\newblock {On the Feasibility of Bell Inequality Violation at ATLAS Experiment
  with Flavor Entanglement of $B^{0}\bar{B}^{0}$ Pairs from $pp$ Collisions}.
\newblock 6 2021.
\newblock arXiv:2106.07399.

\end{thebibliography}
 
\end{document}